\documentclass[useAMS,usenatbib]{mn2e}

\usepackage{revsymb}
\usepackage{amsmath}
\usepackage{amsfonts}
\usepackage{amssymb}
\usepackage{graphicx}

\title[circles-in-the-sky signature]
  {The circles-in-the-sky signature for three spherical universes}
\author[R. Aurich et al.]
  {R.~Aurich,$^1$
  S.~Lustig,$^1$ and F.~Steiner $^1$\\
  $^1$Abteilung Theoretische Physik, Universit\"at Ulm,\\
Albert-Einstein-Allee 11, D-89069 Ulm, Germany}

\date{}

\pagerange{\pageref{firstpage}--\pageref{lastpage}} \pubyear{2005}

\def\LaTeX{L\kern-.36em\raise.3ex\hbox{a}\kern-.15em
    T\kern-.1667em\lower.7ex\hbox{E}\kern-.125emX}

\begin{document}

\def\bfis{\hbox{\scriptsize\rm i}}
\def\bfi{\hbox{\rm i}}
\def\bfj{\hbox{\rm j}}

\newcommand{\apj}{{Astrophys.\ J. }}
\newcommand{\apjs}{{Astrophys.\ J.\ Supp. }}
\newcommand{\apjl}{{Astrophys.\ J.\ Lett. }}
\newcommand{\aj}{{Astron.\ J. }}
\newcommand{\prl}{{Phys.\ Rev.\ Lett. }}
\newcommand{\prd}{{Phys.\ Rev.\ D }}
\newcommand{\mnras}{{Mon.\ Not.\ R.\ Astron.\ Soc. }}
\newcommand{\araa}{{ARA\&A }}
\newcommand{\aap}{{Astron.\ \& Astrophy. }}
\newcommand{\nat}{{Nature }}
\newcommand{\cqg}{{Class.\ Quantum Grav.\ }}

\setlength{\topmargin}{-1cm}

\label{firstpage}

\maketitle

\begin{abstract}
The mysteriously low CMB power on the largest scales might point
to a Universe which consists of a multi-connected space.
In addition to a suppression of large-scale power,
a multi-connected space can be revealed by its
circles-in-the-sky signature.
In this paper, a detailed search for this signature is carried out for those
three homogeneous multi-connected spherical space forms
that lead to the smallest large-scale power.
A simultaneous search for all occurring paired circles
is made using filtered CMB sky maps which enhance the
ordinary Sachs-Wolfe contribution.
A marginal hint is found for the right-handed Poincar\'{e} dodecahedron
at $\Omega_{\hbox{\scriptsize tot}}\simeq 1.015$
and for the right-handed binary tetrahedral space at
$\Omega_{\hbox{\scriptsize tot}}\simeq 1.068$.
However, due to the complicated noise and foreground structure
of the available microwave sky maps,
we cannot draw firm conclusions from our findings.
\end{abstract}

\begin{keywords}
cosmic microwave background - cosmic topology
\end{keywords}


\section{Introduction}

Current observational evidence strongly corroborates the standard
big-bang model based on a Friedmann-Lema\^{\i}tre universe possessing a
space-time structure $\mathbb{R}_+\times{\cal M}$.
Here $\mathbb{R}_+$ describes cosmic time, 
and ${\cal M}$ the three-dimensional comoving space section
of constant curvature $K=+1,0,-1$.
It is well-known that the Einstein field equations describe the local
space-time geometry of the universe,
once the spatial curvature $K$ and the various mass/energy densities
are given, but do not specify its global
spatial geometry, i.\,e.\ its topology.
Thus, in cosmology we are faced with two very fundamental questions:
\begin{itemize}
\item[i)]
What is the value of $K$, i.\,e.\ the curvature of the universe?
\item[ii)]
What is the topology of the three-space ${\cal M}$?
\end{itemize}
While the curvature does constrain the possible topologies,
it does not fix it.
Only if it is {\it assumed} that the space of the universe
is simply-connected, the possible homogeneous three-spaces ${\cal M}$
of constant curvature are given by the three-sphere ${\cal S}^3 (K=+1)$,
Euclidean three-space ${\cal E}^3 (K=0)$, or hyperbolic three-space
${\cal H}^3 (K=-1)$.
At present, there is, however, no compelling theoretical argument
why the spatial geometry of the universe should be simply-connected,
and thus we are confronted with the problem of ``cosmic topology''
(see \cite{Lachieze-Rey_Luminet_1995,Levin_2002} for a review).

The discovery of the temperature fluctuations $\delta T$ of the
cosmic microwave background (CMB) radiation by COBE in 1992
\citep{Smoot_et_al_1992_mnras} and the detailed measurements by WMAP
\citep{Bennett_et_al_2003_mnras} and by other groups offer an
ideal testing ground for a study of cosmic topology.
The first year WMAP data show a slight tendency to a positive curvature
of the universe, at least within the $1\sigma$-band.
In \citet{Aurich_Lustig_Steiner_2005a} we have therefore carried out
a systematic comparison of the WMAP data with the predictions obtained
from a representative sample of homogeneous spherical space forms.
As a result, we have found \citep{Aurich_Lustig_Steiner_2005a}
that only three multi-connected spherical spaces are compatible
with the observed CMB anisotropy.
The most remarkable signature of these spherical universes
is a suppression of the CMB power spectrum at large scales, 
in particular a large suppression of the CMB quadrupole and octopole,
and of the temperature two-point correlation function on large angles
\citep{Luminet_Weeks_Riazuelo_Lehoucq_Uzan_2003,%
Aurich_Lustig_Steiner_2004c,Aurich_Lustig_Steiner_2005a}
in accordance with the COBE and WMAP measurements.
It is worthwhile to emphasize that the suppression on large scales
is a {\it prediction} of these universes and not put in by hand.

Multi-connected spaces can reveal themselves by the so-called
circles-in-the-sky signature \citep{Cornish_Spergel_Starkman_1998b}.
In this paper, we carry out a thorough search of these paired circles
for the above mentioned three spherical spaces.


\section{Spherical Friedmann-Lema\^Itre universes}

In our search for the circles-in-the-sky signature in CMB sky maps,
we are only interested in large-scale fluctuations of the CMB above
horizon at recombination
since these fluctuations should contain the fingerprint of a
non-trivial topology.
The observed relative temperature fluctuations $\delta T/T = O(10^{-5})$
are very small, and thus it is sufficient to use linear perturbation theory.
The four-dimensional space-time metric with scalar perturbations
takes then the simple form (in conformal Newtonian gauge)
\begin{equation}
\label{Eq:Metric}
ds^2 \; = \; a^2(\eta) \, \left[ \, (1+2\Phi) d\eta^2 -
(1-2\Phi) |d\vec x|^2 \, \right]
\hspace{10pt} ,
\end{equation}
where $\eta\in \mathbb{R}_+$ denotes the conformal time and
$a(\eta)$ the cosmic scale factor.
$|d\vec x|$ is the spatial distance on a given spherical space form ${\cal M}$.
For vanishing $\Phi$, $\Phi\equiv 0$, the metric (\ref{Eq:Metric})
is the well-known Friedmann-Robertson-Walker metric.
In writing the metric (\ref{Eq:Metric}) with a non-vanishing
scalar perturbation $\Phi\neq 0$,
we have assumed that the energy-momentum tensor $T_{\mu\nu}$ satisfies
$T_{ij}=0$ for $i\neq j$ $(i,j=1,2,3)$ and that vector and tensor
modes can be neglected in our study of the CMB.

In the following, we shall assume that the total energy density of the
universe is given by a sum of 4 contributions:
radiation, baryonic matter (bar), cold dark matter (cdm), and dark energy,
where the latter is identified with a positive cosmological constant
$\Lambda$.
Then the scale factor $a(\eta)$ is uniquely determined by the
corresponding Friedmann equation.

Let us now discuss the geometry and topology of the spherical space forms.
There are infinitely many spherical spaces which were classified by
\citet{Threlfall_Seifert_1930,Threlfall_Seifert_1932}
(see also \citet{Wolf_1974} and
\citet{Gausmann_Lehoucq_Luminet_Uzan_Weeks_2001})
and are given by the quotient ${\cal M} = {\cal S}^3/\Gamma$
of the three-sphere ${\cal S}^3$ under the action of a discrete
fixed-point free subgroup $\Gamma\subset \hbox{SO}(4)$ of the
isometries of ${\cal S}^3$.
To define the subgroup $\Gamma$,
one makes use of the fact that the unit 3-sphere ${\cal S}^3$
can be identified with the multiplicative, but non-commutative group of
unit quaternions $\{q\}$.
The latter are defined by $q:=w+x\bfi+y\bfj+z\bfi\bfj$,
$(w,x,y,z)\in \mathbb{R}^4$, having unit norm, $|q|^2 = w^2+x^2+y^2+z^2=1$.
Here, the 4 basic quaternions $\{1,\bfi,\bfj,\bfi\bfj\}$ satisfy
the multiplication rules $\bfi^2=\bfj^2=-1$ and $\bfi\bfj=-\bfj\bfi$
plus the property that $\bfi$ and $\bfj$ commute with every real number.
The distance $d(q_1,q_2)$ between two points $q_1$ and $q_2$ on ${\cal S}^3$
is given by $\cos d(q_1,q_2) = w_1w_2+x_1x_2+y_1y_2+z_1z_2$.

The group $\hbox{SO}(4)$ is isomorphic to
${\cal S}^3 \times {\cal S}^3 / \{\pm(1,1)\}$,
and thus the subgroup $\Gamma$ can act on the first ${\cal S}^3$ factor,
on the second  ${\cal S}^3$ factor or on both factors.
Since we are only interested in orientable homogeneous manifolds
${\cal M} = {\cal S}^3/\Gamma$, 
we have to consider either right- or left-handed Clifford translations
$\gamma\in\Gamma$
that act on an arbitrary unit quaternion $q\in{\cal S}^3$ by
left-multiplication, $q \to \gamma q$, respectively,
right-multiplication, $q \to q \gamma$.
The right- (left-) handed Clifford translations act as
right- (left-) handed cork screw translations.

Recently, we have carried out in \citet{Aurich_Lustig_Steiner_2005a}
a systematic comparison of the CMB anisotropy with the WMAP data for
a representative sample of homogeneous spherical spaces
and have found agreement with the data for 3 spaces only:
${\cal T} := {\cal S}^3/T^\star$, ${\cal O} := {\cal S}^3/O^\star$, and
${\cal D} := {\cal S}^3/I^\star$.
Here $T^\star$ denotes the binary tetrahedral group of order 24
generated by the right-handed Clifford translations
$\gamma_1=\bfj$ and $\gamma_2= (1 + \bfi +\bfj +\bfi\bfj)/2$.
$O^\star$ denotes the binary octahedral group of order 48
generated by the right-handed Clifford translations
$\gamma_1=(1+\bfi)/\sqrt 2$ and $\gamma_2= (1 + \bfi +\bfj +\bfi\bfj)/2$.
The last group $I^\star$ corresponding to the
Poincar\'{e} dodecahedron ${\cal D} = {\cal S}^3/I^\star$
is the binary icosahedral group of order 120 and is
generated by the right-handed Clifford translations $\gamma_1=\bfj$ and
$\gamma_2= (\sigma + \frac 1\sigma\bfi + \bfj)/2$ with
$\sigma = (1 + \sqrt 5)/2$.
In addition to these 3 spaces there are the 3 spaces generated by the
analogous left-handed Clifford translations
which implies that altogether we have to consider 6 spherical spaces.

\begin{table*}
\centering
\begin{minipage}{140mm}
\caption{\label{Tab:Spectrum}
The eigenvalue spectrum of the considered spherical spaces ${\cal S}^3$
and ${\cal S}^3/\Gamma$.
}
\begin{tabular}{|c|c|c|}
\hline
${\cal M}$ & wave number spectrum $\{\beta\}$ of manifold ${\cal M}$ &
multiplicity $r^{\cal M}(\beta)$ \\
\hline
${\cal S}^3$ & $\mathbb{N}$  & $\beta^2$ \\
\hline
${\cal S}^3/T^\star$ & $\{1,7,9\} \cup \{2n+1|n\ge 6\}$ &
$\beta\left( 2 \left[\frac{\beta-1}6\right] + \left[\frac{\beta-1}4\right] -
\frac{\beta-3}2\right)$ \\
\hline
${\cal S}^3/O^\star$ & $\{1,9,13,17,19,21\} \cup \{2n+1| n\ge 12\}$ &
$\beta\left( \left[\frac{\beta-1}8\right] + \left[\frac{\beta-1}6\right] +
\left[\frac{\beta-1}4\right] - \frac{\beta-3}2\right)$ \\
\hline
${\cal S}^3/I^\star$ & $\{1,13,21,25,31,33,37,41,43,45,49,51,53,55,57\}$ &
$\beta\left( \left[\frac{\beta-1}{10}\right] + \left[\frac{\beta-1}6\right] +
\left[\frac{\beta-1}4\right] - \frac{\beta-3}2 \right)$ \\
& $\cup \, \{2n+1, n \ge 30\}$ & \\
\hline
\end{tabular}
\end{minipage}
\end{table*}

In order to calculate the CMB fluctuations,
we have to expand the various contributions into the vibrations on the
spaces ${\cal M} = {\cal S}^3/\Gamma$
which are determined by the regular solutions of the Helmholtz equation
\begin{equation}
\label{Eq:Helmholtz}
(\Delta + E_\beta^{\cal M}) \, \psi_\beta^{{\cal M},i}(q) \; = \; 0
\hspace{10pt} , \hspace{10pt}
q \in {\cal M} \; \; ,
\end{equation}
satisfying the periodicity conditions
\begin{equation}
\label{Eq:periodicity_condition}
\psi_\beta^{{\cal M},i}(\gamma_k q) \; = \; \psi_\beta^{{\cal M},i}(q)
\hspace{10pt} , \hspace{10pt}
\forall q \in {\cal M} \; \; , \; \; \forall \gamma_k \in \Gamma
\hspace{10pt} .
\end{equation}
Here $\Delta$ denotes the Laplace-Beltrami operator on ${\cal S}^3$,
and the eigenfunctions $\psi_\beta^{{\cal M},i}(q)$ are in
$L^2({\cal M},d\mu)$ with $d\mu$ the invariant measure on ${\cal S}^3$.
The spectrum on ${\cal M}$ is discrete, and the eigenvalues can be
expressed in terms of the wave number $\beta\in\mathbb{N}$ as
$E_\beta^{\cal M} = \beta^2-1$ and are independent of the
degeneracy index $i=1,\dots,r^{\cal M}(\beta)$,
where $r^{\cal M}(\beta)$ denotes the multiplicity of the mode $\beta$.
Only for ${\cal S}^3$, $\beta$ takes all values in $\mathbb{N}$.
The allowed wave numbers $\beta$ for the above mentioned spherical
manifolds are explicitly known \citep{Ikeda_1995},
see Table \ref{Tab:Spectrum}.

The relative temperature fluctuations of the CMB are caused by
several effects which we shall compute within the tight-coupling
approximation along the lines described in Section 2 of
\citet{Aurich_Lustig_Steiner_Then_2004a}.
The tight-coupling approximation leads to the Sachs-Wolfe formula
(\ref{Eq:Sachs_Wolfe_tight_coupling}),
which incorporates the various effects to be discussed below.
Furthermore, we assume here isentropic initial conditions and for the
primordial power spectrum a scale-invariant Harrison-Zel'dovich spectrum
\begin{equation}
\label{Eq:Harrison_Zeldovich}
P_\Phi(\beta) \; = \; \frac \alpha{\beta(\beta^2-1)}
\hspace{10pt} ,
\end{equation}
where $\alpha$ is a normalization factor
which will be determined from the CMB data
such that the angular power spectrum $\delta T_l^2$
fits the WMAP values in the range $l=20$ to $l=45$.


\section{Cosmic topology and the circles-in-the-sky signature}

There are infinitely many multi-connected spherical space forms
as described in the previous section.
\citet{Aurich_Lustig_Steiner_2004c,Aurich_Lustig_Steiner_2005a}
examined the angular power spectrum
$\delta T_l^2 := l(l+1) C_l / (2\pi)$ and
the temperature two-point correlation function $C( \vartheta )$
for the Poincar\'{e} dodecahedron ${\cal D}$ and for many globally
homogeneous spherical manifolds and
found a good agreement on large scales with the WMAP data only for the
binary tetrahedral, the binary octahedral, and the dodecahedral space forms.
A strong suppression on large scales, as seen in the WMAP data,
is obtained for the binary tetrahedral space form ${\cal T}$
for $\Omega_{\hbox{\scriptsize tot}}=1.06...1.07$,
for the binary octahedral space form ${\cal O}$ for
$\Omega_{\hbox{\scriptsize tot}}=1.03...1.04$,
and for the dodecahedral space form ${\cal D}$ for
$\Omega_{\hbox{\scriptsize tot}}=1.015...1.02$.
This is a hint that one of these topologies could be realized
in our Universe.
Here $\Omega_{\hbox{\scriptsize tot}}:=\varepsilon_{\hbox{\scriptsize tot}}/
\varepsilon_{\hbox{\scriptsize crit}}$ is the energy density parameter,
where $\varepsilon_{\hbox{\scriptsize tot}}$ denotes the total energy density
of the universe and
$\varepsilon_{\hbox{\scriptsize crit}} := \frac{3H_0^2}{8\pi G}$
the critical energy density at the present epoch.

In order to investigate these topologies further,
we perform a special search for the so-called circles-in-the-sky signature
\citep{Cornish_Spergel_Starkman_1998b} in the CMB sky maps.
This topological signal is caused by the multi-connectedness of
the assumed space forms.
The surface of last scattering (SLS) is a 2-sphere with the observer at
the centre.
In multi-connected manifolds, copies of the SLS and the observer
are generated by the periodicity conditions (\ref{Eq:periodicity_condition}).
If the distance between the observer and a copy of him (``mirror observer'')
is smaller than the diameter of the SLS, the two SLS's intersect.
This intersection is a circle.
Both the observer and the copy of the observer see this circle,
but in different directions on the sky.
Because there is really only one observer,
this one observer should detect two circles in different directions
in the microwave sky map.

For a quantitative search of these matched circles,
\citet{Cornish_Spergel_Starkman_1998b} define the quantity
\begin{equation}
\label{sigma_statistik}
S \left(\rho \right) \; := \;
\frac{\langle 2\delta T_a\left(\pm \phi\right)\delta
T_b\left(\phi+\rho \right)\rangle}{\langle\delta T_a^2\left(\phi\right)+
\delta T_b^2\left(\phi\right)\rangle}
\hspace{10pt} .
\end{equation}
$\delta T_a\left(\phi\right)$ and $\delta T_b\left(\phi\right)$ are the
temperature fluctuations along two circles $a$ and $b$ on the microwave sky
having the same radius.
Their centres are in the directions $\hat{n}_a$ and $\hat{n}_b$ on the sky map.
Both circles are parameterized by the angle $\phi$.
The angle $\rho$ gives the relative phase of these circles and $\langle
\hspace{2pt}\rangle :=\int_0^{2\pi}d\phi$.
By searching only for circles that are anti-phased,
which corresponds to the minus sign in the nominator
in (\ref{sigma_statistik}),
one restricts the search to orientable topologies,
but by searching also for circles that are phased,
which corresponds to the plus sign in (\ref{sigma_statistik}),
one can also detect non-orientable topologies.
The $S$ statistic can take values in the interval $[-1,+1]$,
where $S=+1$ corresponds to a perfect agreement between the
temperature fluctuations along the two circles and
$S=-1$ to completely anticorrelated temperature fluctuations.
In the next section, we shall discuss the various effects
which influence the values of the $S$ statistic.

In general, a complete search of matched circles requires a six-dimensional
parameter space:
four parameters for the directions $\hat{n}_a$ and $\hat{n}_b$ of the centres
of the two circles, one parameter for the relative phase $\rho$ of
the two circles, and one parameter for the common radius of the circles.
This has to be done for both orientations of the circles.
Such a search involves a huge computer time.
By searching for matched circles of spherical manifolds,
which are constructed either from right-handed or left-handed
Clifford translations,
we can lower the dimension of the parameter space and
consequently reduce the computational time
\citep{Weeks_Lehoucq_Uzan_2003}.
Since we investigate in this paper three topologies which are globally
homogeneous, we have only to search for back-to-back
matched circles (the circle pairs are separated by $180^{\circ}$)
which reduces the parameter space to four dimensions.
Furthermore, the relative phase $\rho$ of the circles is given.
Tables \ref{paired_circles_dodecahedron}, \ref{paired_circles_octahedron} and
\ref{paired_circles_tetrahedron} display the dependence of the number of
paired circles on the radius $\tau_{\hbox{\scriptsize SLS}}$ of the SLS,
respectively on $\Omega_{\hbox{\scriptsize tot}}$,
for the binary polyhedral spaces $\cal T$, $\cal O$ and $\cal D$.
In addition, the relative phases $\rho$ and the radii
of the paired circles are given.
The column denoted by $\rho$ in these Tables gives the relative phases
of those circles which have to be added to the already present ones
when increasing the value of $\Omega_{\hbox{\scriptsize tot}}$.
The other circles possess the relative phases given in the preceding row.
It is possible to construct these manifolds from right-handed as well as
from left-handed Clifford translations leading to an identical angular power
spectrum $\delta T_l^2$ and the same correlation function $C( \vartheta )$,
but with a reversed sign of $\rho$ in (\ref{sigma_statistik})
which alters the search for matched circles.
Thus, one has to search for the relative phases with
positive and negative sign.
Furthermore, we can restrict our search to circles which are anti-phased,
because the binary polyhedral spaces $\cal T$, $\cal O$ and $\cal D$
are orientable manifolds.

A modification of the quantity $S(\rho)$, equation (\ref{sigma_statistik}),
is introduced by \citet{Cornish_Spergel_Starkman_Komatsu_2003},
but which we do not use here.
Their modified quantity, see  equation (1) in
\citet{Cornish_Spergel_Starkman_Komatsu_2003},
is constructed by expanding the temperature fluctuations
$\delta T_a\left(\phi\right)$ and $\delta T_b\left(\phi\right)$
along the two circles in Fourier series
$\delta T_{a,b}\left(\phi\right) = \sum_m T^{a,b}_m \exp(\bfi m\phi)$.
Then the coefficients $T^{a,b}_m$ are multiplied by a factor $\sqrt{|m|}$
leading to modified temperature fluctuations
$\delta \widetilde{T}_{a,b}\left(\phi\right)$
which in turn are used in (\ref{sigma_statistik}).
This procedure enhances the small scale structure.
As will be discussed below, the available microwave sky maps do not have
sufficient power at small scales, especially near to the first
acoustic peak.
This will be demonstrated by Fig.\,\ref{Fig:CMB_Vergl_ILC_Tegmark}
where the angular power spectrum $\delta T_l^2$ of the CMB
fluctuations is compared with the power contained in the sky maps.
With the multiplication by $\sqrt{|m|}$, this deficit in power is only
remedied in a statistical sense;
the true cosmological signal of the SLS at small scales might be
completely different, of course.
But it is this very cosmological signal one needs for the detection
of paired circles.
Thus, there remains the question about how the procedure of
\citet{Cornish_Spergel_Starkman_Komatsu_2003} affects
the topological signal in true microwave sky maps
being composed of a lot of sky patches.
In the next section, we propose an alternative procedure without
artificially enhancing power at small scales
but instead using a specially designed weight function.

\begin{table*}
\centering
\begin{minipage}{140mm}
\caption{\label{paired_circles_dodecahedron}
The number of paired circles for the Poincar\'{e} dodecahedron  $\cal D$
is given in dependence on the radius $\tau_{\hbox{\scriptsize SLS}}$
of the SLS, respectively on the total density parameter
$\Omega_{\hbox{\scriptsize tot}}$.
The parameters $h=0.70$ and $\Omega_{\hbox{\scriptsize mat}}=0.28$
are held fixed.
Also the relative phases $\rho$ of the circle pairs are shown for the
right-handed (plus sign) and the left-handed (minus sign) Clifford translations
as well as the radii of the circles.}
\begin{tabular}{@{}ccccl@{}}
\hline
$\tau_{\hbox{\scriptsize SLS}}$&$\Omega_{\hbox{\scriptsize tot}}$&
\# of paired circles&$\rho$&radii of the paired circles\\
\hline
$<0.33$&$<1.010$&$0$&-&-\\
$<0.53$&$<1.026$&$6$&$\pm 36^{\circ}$&$<56^{\circ}$\\
$<0.64$&$<1.037$&$16$&$\pm 60^{\circ}$&$<64^{\circ},<39^{\circ}$\\
$<0.79$&$<1.056$&$22$&$\pm 72^{\circ}$&$<71^{\circ},<55^{\circ},<44^{\circ}$\\
$<0.95$&$<1.080$&$37$&$\pm
90^{\circ}$&$<77^{\circ},<65^{\circ},<59^{\circ},<38^{\circ}$\\
$<1.05$&$<1.097$&$43$&$\pm
108^{\circ}$&$<79^{\circ},<71^{\circ},<65^{\circ},<55^{\circ},<38^{\circ}$\\
$<1.26$&$<1.137$&$53$&$\pm
120^{\circ}$&$<84^{\circ},<79^{\circ},<76^{\circ},<71^{\circ},<64^{\circ},<56^{\circ}$\\
$\ge 1.26$&$\ge 1.137$&$59$&$\pm
144^{\circ}$&$\ge84^{\circ},\ge79^{\circ},\ge76^{\circ},\ge71^{\circ},\ge64^{\circ},\ge56^{\circ},\ge9^{\circ}$\\
\hline

\end{tabular}
\end{minipage}
\end{table*}

\begin{table*}
\centering
\begin{minipage}{140mm}
\caption{The same as in Table \ref{paired_circles_dodecahedron} for the binary
octahedral space $\cal O$.\label{paired_circles_octahedron}}
\begin{tabular}{@{}ccccl@{}}
\hline
$\tau_{\hbox{\scriptsize SLS}}$&$\Omega_{\hbox{\scriptsize tot}}$&
\# of paired circles&$\rho$&radii of the paired circles\\
\hline
$<0.40$&$<1.015$&$0$&-&-\\
$<0.53$&$<1.026$&$3$&$\pm 45^{\circ}$&$<45^{\circ}$\\
$<0.79$&$<1.056$&$7$&$\pm 60^{\circ}$&$<66^{\circ},<55^{\circ}$\\
$<1.05$&$<1.097$&$16$&$\pm 90^{\circ}$&$<76^{\circ},<70^{\circ},<55^{\circ}$\\
$<1.18$&$<1.121$&$20$&$\pm
120^{\circ}$&$<80^{\circ},<76^{\circ},<66^{\circ},<44^{\circ}$\\
$\ge 1.18$&$\ge 1.121$&$23$&$\pm
135^{\circ}$&$\ge80^{\circ},\ge76^{\circ},\ge66^{\circ},\ge44^{\circ},\ge6^{\circ}$\\
\hline
\end{tabular}
\end{minipage}
\end{table*}

\begin{table*}
\centering
\begin{minipage}{140mm}
\caption{The same as in Table \ref{paired_circles_dodecahedron} for the binary
tetrahedral space $\cal T$.\label{paired_circles_tetrahedron}}
\begin{tabular}{@{}ccccl@{}}
\hline
$\tau_{\hbox{\scriptsize SLS}}$&$\Omega_{\hbox{\scriptsize tot}}$&
\# of paired circles&$\rho$&radii of the paired circles\\
\hline
$<0.53$&$<1.026$&$0$&-&-\\
$<0.79$&$<1.056$&$4$&$\pm 60^{\circ}$&$<55^{\circ}$\\
$<1.05$&$<1.097$&$7$&$\pm 90^{\circ}$&$<71^{\circ},<55^{\circ}$\\
$\ge 1.05$&$\ge 1.097$&$11$&$\pm
120^{\circ}$&$\ge71^{\circ},\ge55^{\circ},\ge7^{\circ}$\\
\hline
\end{tabular}
\end{minipage}
\end{table*}

One can perform a very specific search by using the information about
the group $\Gamma$ which defines the manifold ${\cal S}^3/\Gamma$.
Fixing the radius $\tau_{\hbox{\scriptsize SLS}}$ of the SLS in the case of
a homogeneous spherical manifold, the number of paired circles,
their relative phases $\rho$ and their relative orientation are determined.
Only a common rotation of the whole configuration of the circles
is left free which can be parameterised by three Euler angles.
This fact can be used to carry out a simultaneous search over all
matched circles by defining the estimator
\begin{equation}
\label{sigma_n_statistik}
\Sigma \left(n \right) \; := \;
\hbox{sign}\left(\sum_{i=1}^{n}S_i \cdot \left|S_i\right|\right)
\cdot \sqrt{\frac{\left|\sum_{i=1}^{n}S_i \cdot \left|S_i\right|\right|}{n}}
\hspace{10pt} .
\end{equation}
$S_{i}$ is the $S$ statistic of the $i$-th paired circles and
$n$ is their total number.
$\Sigma$ varies between $-1$ and $+1$.
This search has a four-dimensional parameter space: three parameters for all
possible orientations of the fundamental cell and one parameter for the
radius $\tau_{\hbox{\scriptsize SLS}}$ of the SLS.
The $\Sigma$ estimator gives a better signal than the $S$ statistic, if there
are genuine matched circles; furthermore, it should be possible to detect
with $\Sigma$ smaller paired circles.
Such a special search cannot be used to check all possible topologies,
but if there is a hint towards a special topology,
like the binary polyhedral spaces $\cal T$, $\cal O$ and $\cal D$,
it can be carried out.


\section{Search for matched circles in the WMAP sky maps}

In order to perform the search for matched circles in actual sky maps,
one has to take several points into account.
The most important one is that one does not observe
the pure signal of the temperature fluctuations on the SLS
proportional to the scalar perturbation $\Phi$,
which displays as a scalar quantity a clear sign of the topology.
Even ignoring the detector noise, foreground sources and all but one
secondary contribution,
i.\,e.\ not excluding the integrated Sachs-Wolfe effect,
one has three main contributions as revealed by the
Sachs-Wolfe formula $(\tau(\eta):=\eta_0 - \eta)$
\begin{eqnarray} \nonumber
\frac{\delta T}T(\hat n) & = &
{\sum_{\beta\ge 3}} '\; \sum_{i=1}^{r^{\cal M}(\beta)} \;
\\ & &  \nonumber \hspace{-50pt}
\left[ \left( \Phi_\beta^i(\eta) +
\frac{\delta_{\gamma,\beta}^i(\eta)}4 +
\frac{a(\eta) V_{\gamma,\beta}^i(\eta)}{E_\beta}
\frac{\partial}{\partial \tau} \right)
\Psi_\beta^{{\cal M},i}(\tau(\eta),\theta,\phi)
\right]_{\hspace*{-10pt}\eta=\eta_{\hbox{\scriptsize{SLS}}}}
\\ & & \hspace{-50pt}
\label{Eq:Sachs_Wolfe_tight_coupling}
\, + \,
2 \; {\sum_{\beta\ge 3}} '\; \sum_{i=1}^{r^{\cal M}(\beta)} \;
\int_{\eta_{\hbox{\scriptsize{SLS}}}}^{\eta_0} d\eta \,
\frac{\partial\Phi_\beta^i(\eta)}{\partial\eta} \,
\Psi_\beta^{{\cal M},i}(\tau(\eta),\theta,\phi)
\hspace{5pt} .
\end{eqnarray}
Here the summation runs only over those modes which exist for
a given manifold, see Table \ref{Tab:Spectrum}.
The first two terms in (\ref{Eq:Sachs_Wolfe_tight_coupling}),
determined by the expansion coefficients $\Phi_\beta^i(\eta)$
of the gravitational potential $\Phi$
and the intrinsic perturbation in the radiation density
$\delta_{\gamma,\beta}^i(\eta)$,
yield the effective temperature perturbation on the SLS,
i.\,e.\ the ordinary Sachs-Wolfe (SW) contribution
which encodes the desired topological information.
The next term involving the spatial covariant divergence of the
velocity field $V_{\gamma,\beta}^i(\eta)$ is the Doppler contribution
which spoils the topological signal
because the projection of the velocities in the directions of the
observer and the ``mirror'' observer are not the same.
The last term, the integral over the photon path, yields the
integrated Sachs-Wolfe contribution (ISW).
Since the photon paths towards the observer and the mirror observer
are not identified, this contribution also degrades the
topological signal.

\begin{figure}
\begin{center}
\vspace*{-90pt}
\begin{minipage}{11cm}
\hspace*{-65pt}\includegraphics[width=11.0cm]{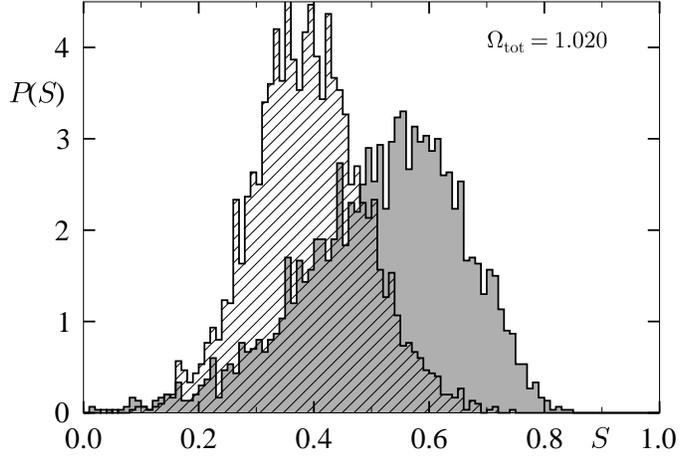}
\end{minipage}
\end{center}
\vspace*{-50pt}
\caption{\label{Fig:Sigma_S3_Dode}
The distribution of the $S$ values for matched pairs of circles
is shown for the dodecahedral space ${\cal D}$ as the grey histogram.
For comparison, the distribution for $S_{\hbox{\scriptsize max}}$ for the
simply-connected space ${\cal S}^3$ is shown as a shaded histogram,
where the positions of the pairs are assumed to be the same as
in the dodecahedral case.
The circles have a radius of $49.8^\circ$ determined by
$\Omega_{\hbox{\scriptsize tot}}=1.020$.
The other cosmological parameters are $h=0.70$,
$\Omega_{\hbox{\scriptsize mat}}=0.28$ and
$\Omega_{\hbox{\scriptsize bar}}=0.046$.
}
\end{figure}

Fig.\ \ref{Fig:Sigma_S3_Dode} demonstrates these degrading effects.
The grey histogram shows the distribution of $S(\rho)$
(see Eq.~(\ref{sigma_statistik}))
with $\rho=36^\circ$ for 3000 pairs of matched circles obtained
from simulations of sky maps for the dodecahedral space ${\cal D}$
for a fixed set of cosmological parameters.
Without degrading effects, one would observe a delta-like peak at $S=1$.
As revealed by the grey histogram, the Doppler and ISW contributions
lead to a broad distribution with a maximum around $S=0.6$.
The shaded histogram shows the distribution of
$S_{\hbox{\scriptsize max}} = \max_\rho\{S(\rho)\}$
for 3000 pairs of circles of simulations of
sky maps for the simply connected space ${\cal S}^3$.
(For the positions of the circles we use the positions occurring
in the dodecahedral simulations.)
Since this space has no matched pairs at all, the shaded histogram
provides the null hypothesis for the detection of a
topological signature.

Assume for a moment that the sky map has a genuine circle signature.
Then one has to scan the map for all possible orientations of
the pairs of circles.
The wrong guesses should lead to a distribution similar to that
of the simply-connected ${\cal S}^3$, 
whereas the few correct guesses are distributed as in the
grey histogram.
With the parameters chosen in Fig.\ \ref{Fig:Sigma_S3_Dode},
the dodecahedral space ${\cal D}$ has only six pairs
(see Table \ref{paired_circles_dodecahedron}).
Compared with the huge number of wrong guesses distributed as
the shaded histogram,
the topological signature can easily be washed out.

In order to enhance the topological signal, we apply a special
weight function to the sky maps before carrying out
the search for matched pairs of circles.
Furthermore, we carry out the search for all circles of a given
topology simultaneously, see equation (\ref{sigma_n_statistik}),
which should lead to a larger signal due to the correlation.

\begin{figure}
\begin{center}
\vspace*{-90pt}
\begin{minipage}{11cm}
\hspace*{-65pt}\includegraphics[width=11.0cm]{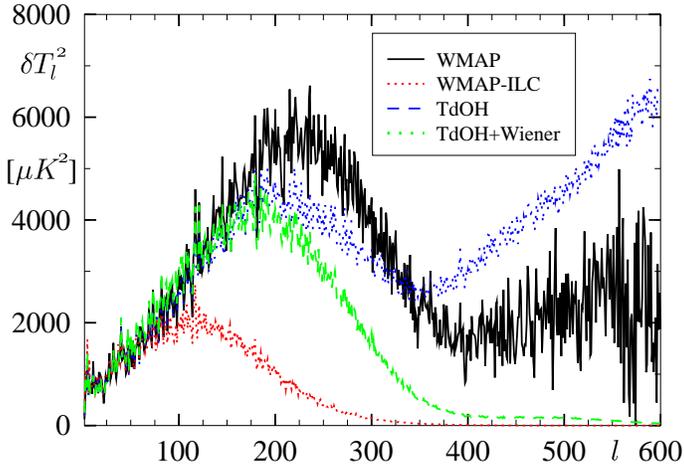}
\end{minipage}
\end{center}
\vspace*{-50pt}
\caption{\label{Fig:CMB_Vergl_ILC_Tegmark}
The angular power spectrum $\delta T_l^2$
of the WMAP analysis (black) of the first year data is shown .
In addition the $\delta T_l^2$ spectra are shown obtained from
the ILC map (red), the TdOH map without (blue) and with (green)
a Wiener filter applied.
}
\end{figure}

Before discussing our weight function,
we describe which angular scales seem to be preferable for such a search.
Only maps which cover the complete sky can be used,
which restricts us to the WMAP observations.
In the following we concentrate on three sky maps derived
from the WMAP observation:
the ILC map of \citet{Bennett_et_al_2003_mnras} and the two maps of
\cite{Tegmark_deOliveira_Costa_Hamilton_2003} with and without
the application of a Wiener filter, respectively.
The quality of these maps is discussed in
\citep{Tegmark_deOliveira_Costa_Hamilton_2003,Efstathiou_2004,%
Eriksen_Banday_Gorski_Lilje_2004,Eriksen_Banday_Gorski_Lilje_2005}.
In order to reveal the angular scales at which these maps
can be used for a circle search, we compare the angular power spectra
$\delta T_l^2$ obtained from these maps using HEALPix
(HEALPix web-site: http://healpix.jpl.nasa.gov)
with the angular power spectra obtained from a direct analysis
of the WMAP first year data \citep{Bennett_et_al_2003_mnras}.
All four angular power spectra are shown in Fig.\
\ref{Fig:CMB_Vergl_ILC_Tegmark}.
Although it would be preferable to carry out the circle search
near to the first acoustic peak,
Fig.\ \ref{Fig:CMB_Vergl_ILC_Tegmark} reveals
that the three derived maps do not have angular power spectra
matching the first acoustic peak as observed in the WMAP spectrum.
Only at scales larger than the first acoustic peak $(l \lesssim 100)$
do the four spectra agree.
Thus we cannot search for the topological signature on angular scales
corresponding to the first acoustic peak.
Instead, we introduce our advertised weight function
which takes into account only the scales with $l\leq 70$.

\begin{figure}
\begin{center}
\vspace*{-90pt}
\begin{minipage}{11cm}
\hspace*{-65pt}\includegraphics[width=11.0cm]{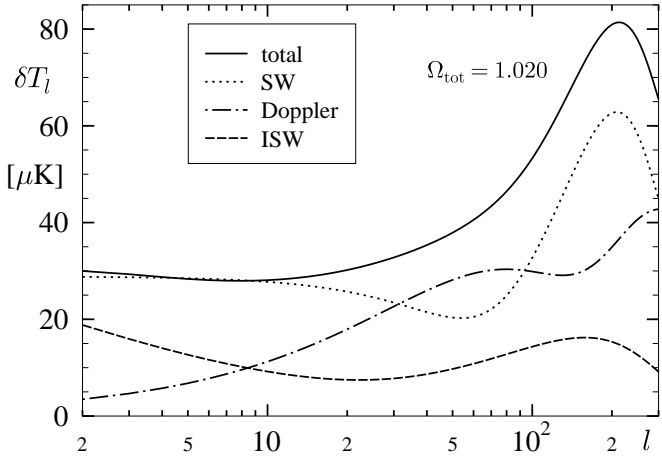}
\end{minipage}
\end{center}
\vspace*{-50pt}
\caption{\label{Fig:delta_T_contributions_S3}
The angular power spectrum $\delta T_l$ is shown for an ${\cal S}^3$ model
with $\Omega_{\hbox{\scriptsize tot}}=1.020$
together with the ordinary Sachs-Wolfe (SW),
the Doppler, and the integrated Sachs-Wolfe (ISW) contribution.
}
\end{figure}

\begin{figure}
\begin{center}
\vspace*{-90pt}
\begin{minipage}{11cm}
\hspace*{-65pt}\includegraphics[width=11.0cm]{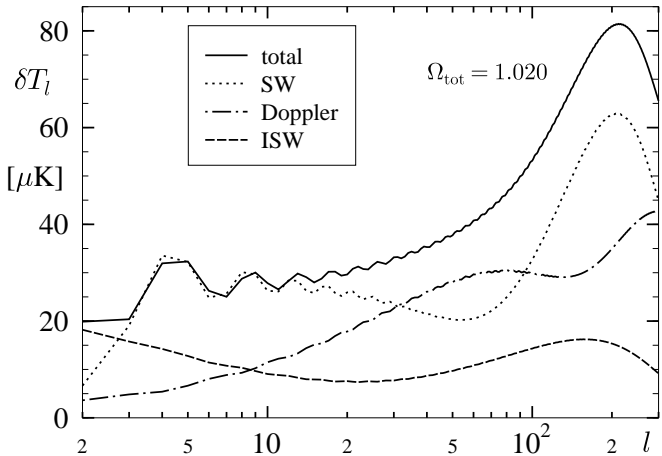}
\end{minipage}
\end{center}
\vspace*{-50pt}
\caption{\label{Fig:delta_T_contributions_Dod}
The same quantities are shown as in
Fig.~\ref{Fig:delta_T_contributions_S3},
but for the Poincar\'e dodecahedral space ${\cal D}$.
}
\end{figure}

The mean value of the angular power spectrum
$\delta T_l :=(l(l+1) C_l/(2\pi))^{1/2}$
is shown in Figs.~\ref{Fig:delta_T_contributions_S3} and
\ref{Fig:delta_T_contributions_Dod}
for the ${\cal S}^3$ model and for
the Poincar\'e dodecahedral space ${\cal D}$, respectively.
We also show the individual contributions to the total
temperature fluctuations as discussed below
equation (\ref{Eq:Sachs_Wolfe_tight_coupling}),
i.\,e.\ the ordinary Sachs-Wolfe contribution (SW)
encoding the topological signature,
and the Doppler and the integrated Sachs-Wolfe (ISW) contributions,
which both degrade the topological signal.
The topological signal should be well preserved in those
$l$-ranges, where the SW contribution dominates the other two,
i.\,e.\ in the large scale $l \lesssim 30$ region and
the region around the first acoustic peak with
$100 \lesssim l \lesssim 300$.
However, as revealed by Fig.\ \ref{Fig:CMB_Vergl_ILC_Tegmark},
neither the ILC map nor one of the TdOH maps have enough power
up to $l \simeq 300$ in order to give a reliable map for
the temperature fluctuations around the first acoustic peak.

\begin{figure}
\begin{center}
\vspace*{-90pt}
\begin{minipage}{11cm}
\hspace*{-65pt}\includegraphics[width=11.0cm]{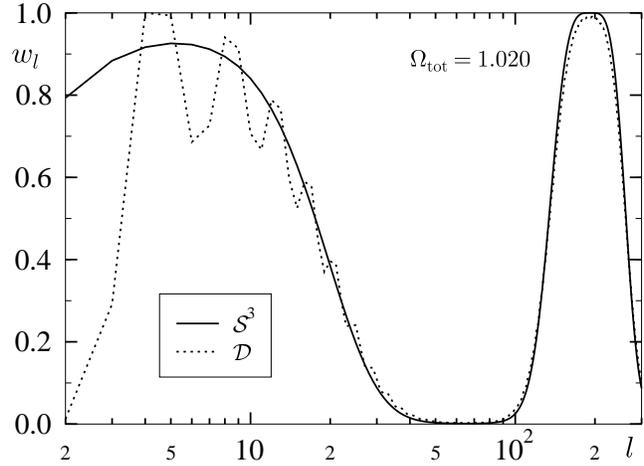}
\end{minipage}
\end{center}
\vspace*{-50pt}
\caption{\label{Fig:weight_function}
The weight function $w_l$ is shown for the ${\cal S}^3$ space (full curve)
and for the Poincar\'e dodecahedral space ${\cal D}$ (dotted curve).
In this figure, the cut-off is $l_{\hbox{\scriptsize cut}}=300$ and not
$l_{\hbox{\scriptsize cut}}=70$ as it is set in the following
computations.
}
\end{figure}

In order to enhance the topological signal,
we decompose the sky maps into the expansion coefficients $a_{lm}$
with respect to spherical harmonics $Y_{lm}(\hat n)$,
multiply the coefficients $a_{lm}$ by weight coefficients $w_l$,
and generate from them a new map.
For the weight function $w_l$,
we choose for $l=2,\dots,l_{\hbox{\scriptsize cut}}$
\begin{equation}
\label{Eq:weight_function}
w_l \; := \;
\exp\left\{-(\delta T_{\hbox{\scriptsize max}} -
\delta T^w_l )^2 \, / \, \delta T_{\hbox{\scriptsize max}}^2 \right\}
\end{equation}
with
$$
\delta T^w_l \; := \;
\delta T_l^{\hbox{\scriptsize SW}} \, - \, \left(
\delta T_l^{\hbox{\scriptsize Doppler}} + \delta T_l^{\hbox{\scriptsize ISW}}
\right)
$$
and
$$
\delta T_{\hbox{\scriptsize max}} \; := \;
\max\left\{ \delta T^w_l \, , \, l=2,\dots,l_{\hbox{\scriptsize cut}} \right\}
\hspace{10pt} .
$$
For $l>l_{\hbox{\scriptsize cut}}$, we set $w_l=0$.
The coefficients $\delta T_l^{\hbox{\scriptsize NSW}}$,
$\delta T_l^{\hbox{\scriptsize Doppler}}$ and
$\delta T_l^{\hbox{\scriptsize ISW}}$ are computed using the
corresponding terms in equation (\ref{Eq:Sachs_Wolfe_tight_coupling}).
This weight function $w_l$ is unity for the $l$ value with the
largest SW contribution relative to the other two contributions
with $l\leq l_{\hbox{\scriptsize cut}}$.
The modes are the more suppressed the more important the Doppler
and the ISW contributions are.
In Fig. \ref{Fig:weight_function} we show an example for the
Poincar\'e dodecahedral space ${\cal D}$ with the
cosmological parameters of Fig. \ref{Fig:Sigma_S3_Dode}.
A significant contribution comes from large scale modes with $l \lesssim 20$.
The strong decline of $w_l$ from $l=20$ to $l=40$ is mainly due to the
growing Doppler term (see Figs.\ \ref{Fig:delta_T_contributions_S3} and
\ref{Fig:delta_T_contributions_Dod}).
Around $l=60..70$, the $w_l$'s are close to zero because the
Doppler contribution is larger than the SW contribution.
Thus, we avoid a Gibb's like phenomenon by giving
$l_{\hbox{\scriptsize cut}}$ a value just within this region.
In the following analysis, we use $l_{\hbox{\scriptsize cut}}=70$.
The contributions from the acoustic peak would also be important
as revealed by Fig.\ \ref{Fig:weight_function},
however, as discussed above, no reliable maps up to $l\simeq 300$ are at hand.

The weight function $w_l$ differs significantly
for the simply-connected space ${\cal S}^3$ and
a multi-connected space like ${\cal D}$ at small values of $l$.
This is due to the different eigenvalue spectra of these systems
(see Table \ref{Tab:Spectrum}).
The modes $l=2$ and $l=3$ are strongly suppressed compared to
the simply-connected case
being a welcomed property in view of the current discussion
of the validity of the maps with respect to the lowest modes
\citep{Schwarz_Starkman_Huterer_Copi_2004}.
For higher values of $l$, the weight function $w_l$ of multi-connected spaces
fluctuates about the values for the simply-connected case.

\begin{figure}
\begin{center}
\vspace*{-90pt}
\begin{minipage}{11cm}
\hspace*{-65pt}\includegraphics[width=11.0cm]{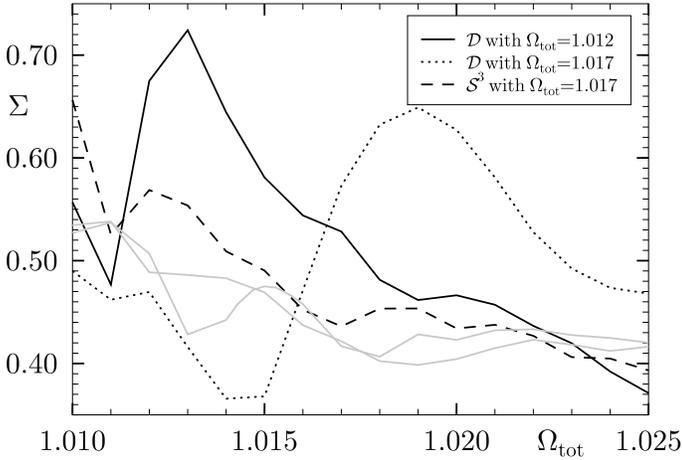}
\end{minipage}
\end{center}
\vspace*{-50pt}
\caption{\label{Fig:Sigma_Dodekaeder_Simulation}
The $\Sigma$ estimator is shown for two simulated sky maps for
the Poincar\'e dodecahedral space ${\cal D}$ and for
the simply connected ${\cal S}^3$.
The light grey curves show the corresponding values obtained
from the Wiener filtered TdOH map
for the left- and right-handed dodecahedral spaces ${\cal D}$.
}
\end{figure}

In the following analysis, we use the specially designed $\Sigma$ measure
with respect to the space form for which the topological signal is looked for.
This means that in the case of the Poincar\'e dodecahedral space ${\cal D}$,
the weight function $w_l$ is computed from the eigenvalues of ${\cal D}$,
and all circles are used simultaneously occurring for that space
for the chosen value of $\Omega_{\hbox{\scriptsize tot}}$.
This is the best method applied up to now for the search of a
given multi-connected space form.
In Fig.~\ref{Fig:Sigma_Dodekaeder_Simulation} we show the results
for two sky simulations for the dodecahedral space ${\cal D}$
where all modes up to $\beta=185$ have been included
which should suffice to obtain the contributions required
by the weight function.
One observes pronounced maxima near the values of
$\Omega_{\hbox{\scriptsize tot}}$ which are selected for the simulation.
In addition, the result for a simulation of the simply-connected ${\cal S}^3$
is shown displaying no large maxima.
For very small circles one can by chance find a better match
than for larger ones.
Thus, as a general trend, one observes a decline of $\Sigma$ with
increasing values of $\Omega_{\hbox{\scriptsize tot}}$
since then the radii of the circles increase.
This behaviour is revealed by the ${\cal S}^3$ simulation
which has, of course, no circles at all.
As grey curves the results belonging to the Wiener filtered TdOH map are shown
in Fig.~\ref{Fig:Sigma_Dodekaeder_Simulation}, see below.
Fig.~\ref{Fig:Sigma_Oktaeder_Simulation} compares the $\Sigma$ estimator
for the octahedral space ${\cal O}$ with the one obtained from
a simply-connected ${\cal S}^3$ space.
The result for the Wiener filtered TdOH map is also shown.
This demonstrates that our $\Sigma$ measure yields a sufficiently
large topological signal despite the ``contaminating'' Doppler and
ISW contributions.

\begin{figure}
\begin{center}
\vspace*{-90pt}
\begin{minipage}{11cm}
\hspace*{-65pt}\includegraphics[width=11.0cm]{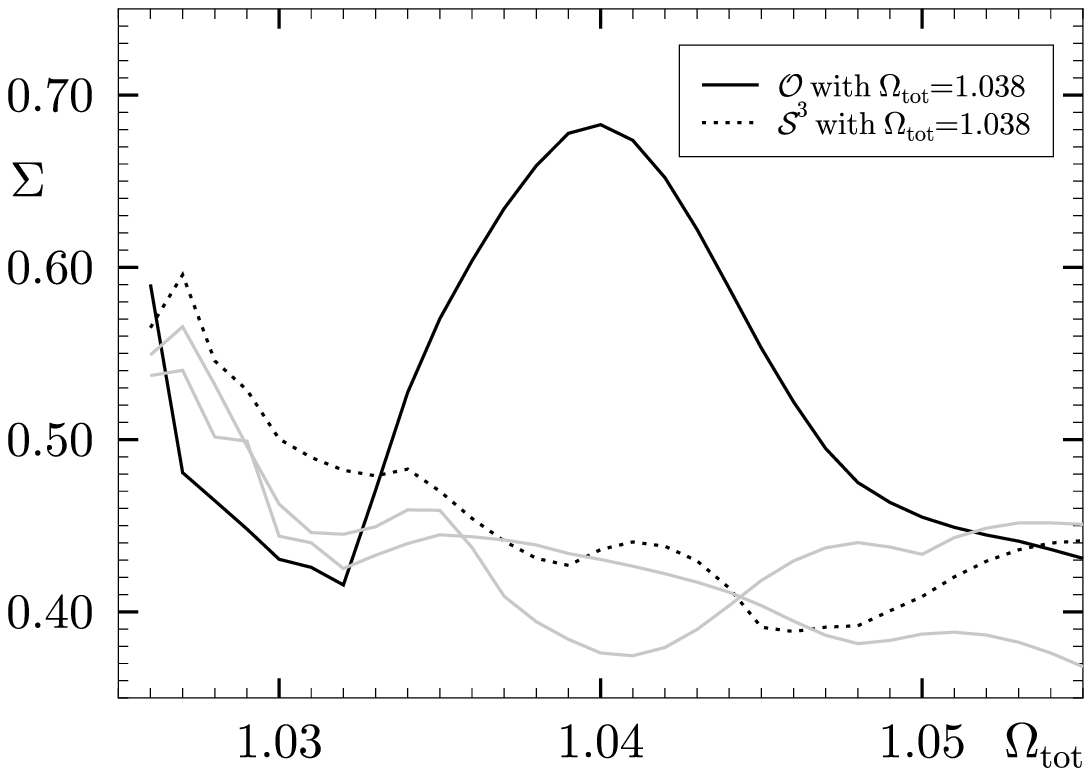}
\end{minipage}
\end{center}
\vspace*{-50pt}
\caption{\label{Fig:Sigma_Oktaeder_Simulation}
The $\Sigma$ estimator is shown obtained from simulated sky maps for
the binary octahedral space ${\cal O}$ and for
the simply connected ${\cal S}^3$.
The light grey curves show the corresponding values obtained
from the Wiener filtered TdOH map
for the left- and right-handed binary octahedral spaces ${\cal O}$.
}
\end{figure}

\begin{figure}
\begin{center}
\vspace*{-90pt}
\begin{minipage}{11cm}
\hspace*{-65pt}\includegraphics[width=11.0cm]{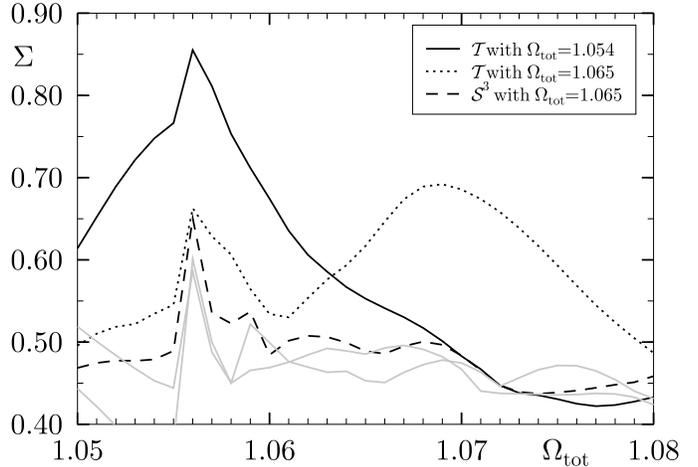}
\end{minipage}
\end{center}
\vspace*{-50pt}
\caption{\label{Fig:Sigma_Tetraeder_Simulation}
The $\Sigma$ estimator is shown for two simulated sky maps for
the binary tetrahedral space ${\cal T}$ and for
the simply connected ${\cal S}^3$.
The light grey curves show the corresponding values obtained
from the Wiener filtered TdOH map
for the left- and right-handed binary tetrahedral spaces ${\cal T}$.
}
\end{figure}

In the considered interval of $\Omega_{\hbox{\scriptsize tot}}$,
the Poincar\'e dodecahedral space ${\cal D}$ produces six matched pairs of
circles on the sky (see Table \ref{paired_circles_dodecahedron}).
However, with increasing values of $\Omega_{\hbox{\scriptsize tot}}$,
the number of pairs of circles increases until the maximal possible
number of 59 is reached.
In general, the maximal number of paired circles for a manifold
${\cal S}^3/\Gamma$ is $\frac N2-1$, where $N$ is the order of the group.
At those values of $\Omega_{\hbox{\scriptsize tot}}$,
where the number of pairs of circles makes a jump,
the $\Sigma$ estimator jumps accordingly to larger values,
which is thus not a sign of the detection of a
multi-connected space form, but simply an artefact of the
increasing contributing number of pairs to the $\Sigma$ estimator.
This is the case for the binary tetrahedral space ${\cal T}$
at $\Omega_{\hbox{\scriptsize tot}} = 1.056$
where the number of circle pairs jumps from 4 to 7
(see Table \ref{paired_circles_tetrahedron}).
In Fig.~\ref{Fig:Sigma_Tetraeder_Simulation}
the $\Sigma$ estimator is shown for two sky simulations for the
binary tetrahedral space ${\cal T}$ having a value of
$\Omega_{\hbox{\scriptsize tot}} = 1.054$ and
$\Omega_{\hbox{\scriptsize tot}} = 1.065$, respectively.
Both show a maximum near to the expected value of
$\Omega_{\hbox{\scriptsize tot}}$,
but in addition a superimposed peak at
$\Omega_{\hbox{\scriptsize tot}} = 1.056$ is visible
where the number of circle pairs jumps from 4 to 7.
The same peak is seen in the ${\cal S}^3$ simulation
which has no matched pairs of circles.
As in Fig.~\ref{Fig:Sigma_Dodekaeder_Simulation},
the grey curves display the results belonging to the Wiener filtered TdOH map.
(The peak at $\Omega_{\hbox{\scriptsize tot}} = 1.056$ is also seen in
Figs.~\ref{Fig:Sigma_WMAP_TdOH_Tetraeder_rh} and
\ref{Fig:Sigma_WMAP_TdOH_Tetraeder_lh}.)


Now, we apply the $\Sigma$ estimator to the 
Poincar\'e dodecahedral space ${\cal D}$,
the binary octahedral space ${\cal O}$, and
the binary tetrahedral space ${\cal T}$
using three sky maps obtained from the WMAP measurements
\citep{Bennett_et_al_2003_mnras}.
The maps are constructed using different combinations
of the data obtained in different frequency bands
in order to suppress foreground sources.
Different combinations are used for different sky patches.
This is not ideal for the search of pairs of circles,
since this leads to an inhomogeneous structure in the maps.
The $\Sigma$ estimator correlates the values on the circles
which are then obtained by different combinations of the
original data.
The WMAP team explicitly states that their ILC map should not
be used for cosmological studies.
But due to a lack of an alternative, we apply,
with this drawback in mind, the $\Sigma$ estimator to the
ILC map produced by the WMAP team \citep{Bennett_et_al_2003_mnras},
and to the two maps of \cite{Tegmark_deOliveira_Costa_Hamilton_2003}.

\begin{figure}
\begin{center}
\vspace*{-90pt}
\begin{minipage}{11cm}
\hspace*{-65pt}\includegraphics[width=11.0cm]{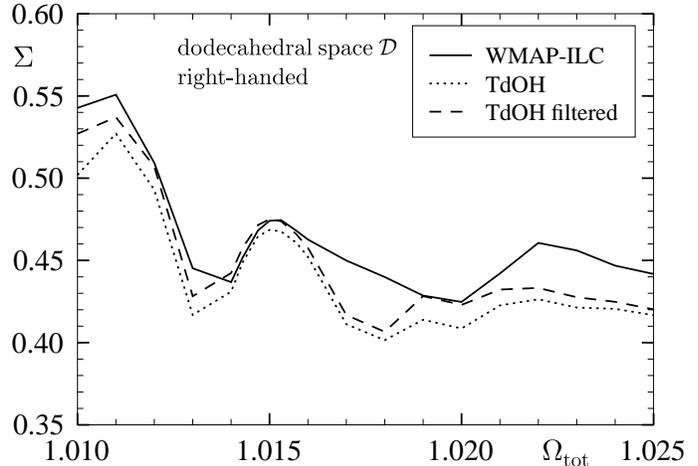}
\end{minipage}
\end{center}
\vspace*{-50pt}
\caption{\label{Fig:Sigma_WMAP_TdOH_Dodekaeder_rh}
The results for the $\Sigma$ estimator are shown evaluated for the
right-handed Poincar\'e dodecahedral space ${\cal D}$ for
the ILC map and the two maps of
\protect\cite{Tegmark_deOliveira_Costa_Hamilton_2003}.
}
\end{figure}

\begin{figure}
\begin{center}
\vspace*{-90pt}
\begin{minipage}{11cm}
\hspace*{-65pt}\includegraphics[width=11.0cm]{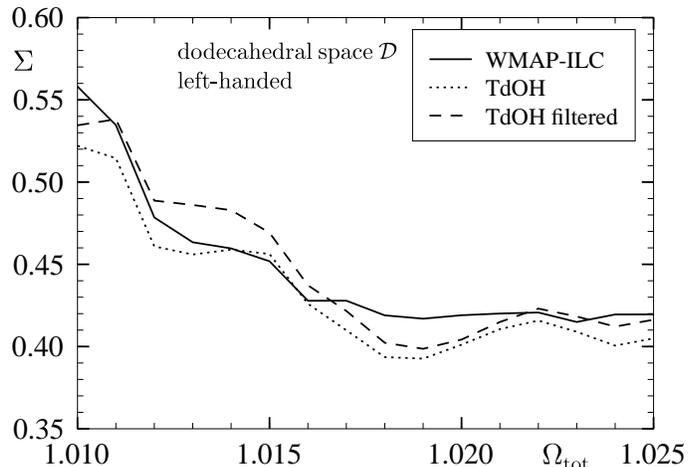}
\end{minipage}
\end{center}
\vspace*{-50pt}
\caption{\label{Fig:Sigma_WMAP_TdOH_Dodekaeder_lh}
The same as in Fig.~\ref{Fig:Sigma_WMAP_TdOH_Dodekaeder_rh}
is shown for the left-handed Poincar\'e dodecahedral space ${\cal D}$.
}
\end{figure}

\begin{figure}
\begin{center}
\vspace*{-90pt}
\begin{minipage}{11cm}
\hspace*{-65pt}\includegraphics[width=11.0cm]{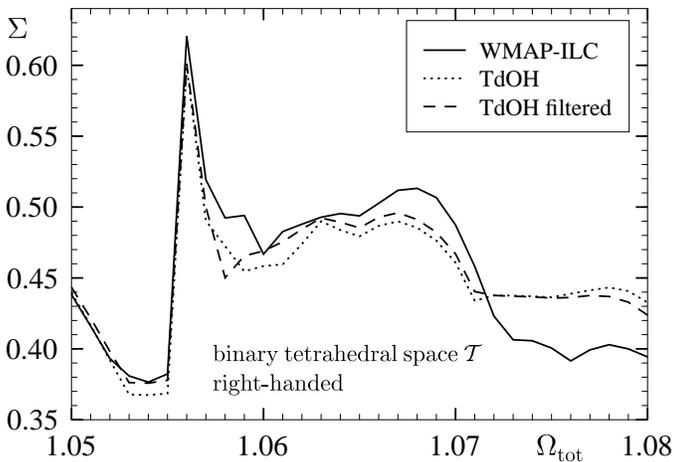}
\end{minipage}
\end{center}
\vspace*{-50pt}
\caption{\label{Fig:Sigma_WMAP_TdOH_Tetraeder_rh}
The same as in Fig.~\ref{Fig:Sigma_WMAP_TdOH_Dodekaeder_rh}
is shown for the right-handed binary tetrahedral space ${\cal T}$.
}
\end{figure}

\begin{figure}
\begin{center}
\vspace*{-90pt}
\begin{minipage}{11cm}
\hspace*{-65pt}\includegraphics[width=11.0cm]{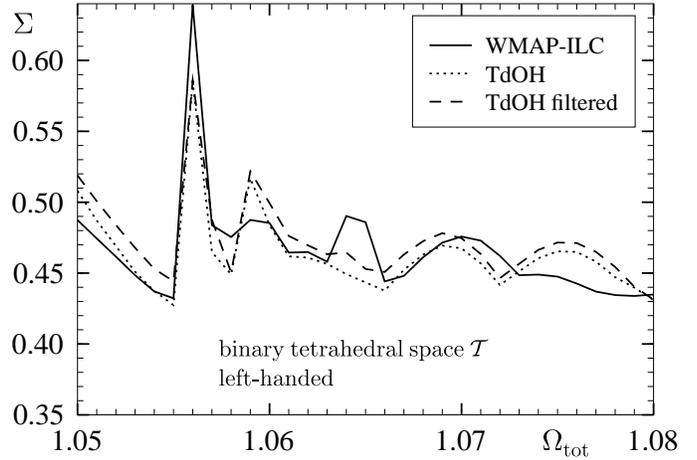}
\end{minipage}
\end{center}
\vspace*{-50pt}
\caption{\label{Fig:Sigma_WMAP_TdOH_Tetraeder_lh}
The same as in Fig.~\ref{Fig:Sigma_WMAP_TdOH_Dodekaeder_rh}
is shown for the left-handed binary tetrahedral space ${\cal T}$.
}
\end{figure}

In Figs.~\ref{Fig:Sigma_WMAP_TdOH_Dodekaeder_rh} and
\ref{Fig:Sigma_WMAP_TdOH_Dodekaeder_lh},
the results for the $\Sigma$ estimator are shown for the
Poincar\'e dodecahedral space ${\cal D}$
for the right- and left-handed space form, respectively.
For the dodecahedral space ${\cal D}$,
the best match to the angular power spectrum measured by WMAP
is obtained around $\Omega_{\hbox{\scriptsize tot}}=1.018$
in \citet{Aurich_Lustig_Steiner_2005a}.
Is there something special in Figs.~\ref{Fig:Sigma_WMAP_TdOH_Dodekaeder_rh}
and \ref{Fig:Sigma_WMAP_TdOH_Dodekaeder_lh} near to this value of
$\Omega_{\hbox{\scriptsize tot}}$?
Whereas the left-handed variant shown in
Fig.~\ref{Fig:Sigma_WMAP_TdOH_Dodekaeder_lh}
displays only the smooth decline of $\Sigma$ due to the
increasing size of the circles with increasing values of
$\Omega_{\hbox{\scriptsize tot}}$,
the right-handed space form shown in
Fig.~\ref{Fig:Sigma_WMAP_TdOH_Dodekaeder_rh}
displays a clear maximum around $\Omega_{\hbox{\scriptsize tot}}=1.015$
in all 3 maps which is particularly distinct in the two maps
of \cite{Tegmark_deOliveira_Costa_Hamilton_2003}.
One is tempted to interpret the peak at $\Omega_{\hbox{\scriptsize tot}}=1.015$
as an indication that the shape of the universe is that of the
right-handed Poincar\'e dodecahedron ${\cal D}$.
However, the maximum seen in Fig.~\ref{Fig:Sigma_WMAP_TdOH_Dodekaeder_rh}
is far less pronounced than the maxima
obtained from our simulations of sky maps for the
dodecahedral space ${\cal D}$, as can be seen by a comparison
with Fig.~\ref{Fig:Sigma_Dodekaeder_Simulation}
where the simulations (full and dotted curves) are presented
for two different $\Omega_{\hbox{\scriptsize tot}}$-values
together with the Wiener filtered TdOH values (grey curves).
There, one observes that the height of the maximum is of the same order
as the amplitude of the ${\cal S}^3$ simulation
possessing no matched circle pairs.
However, these simulations are not based on a patchwork of sky maps,
and it is expected that such a patchwork degrades the signal
for a multi-connected space.
But since its is unknown by how much the signal is degraded,
we cannot draw firm conclusions from this structure at this moment
due to the limited quality of present CMB data.

A possible hint for the left-handed dodecahedral space ${\cal D}$
is found by \citet{Roukema_et_al_2004} who find a slight signal
for six circle pairs with $11\pm 1^\circ$ in the ILC map
which is not revealed distinctly in
Fig.\ \ref{Fig:Sigma_WMAP_TdOH_Dodekaeder_lh}.

\begin{figure}
\begin{center}
\vspace*{-90pt}
\begin{minipage}{11cm}
\hspace*{-65pt}\includegraphics[width=11.0cm]{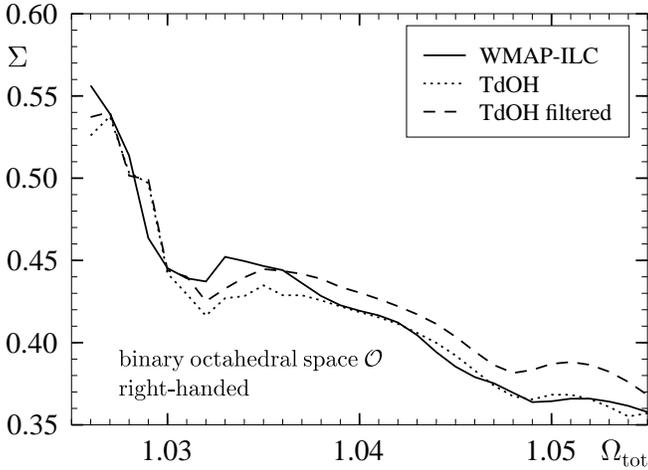}
\end{minipage}
\end{center}
\vspace*{-50pt}
\caption{\label{Fig:Sigma_WMAP_TdOH_Oktaeder_rh}
The same as in Fig.~\ref{Fig:Sigma_WMAP_TdOH_Dodekaeder_rh}
is shown for the right-handed binary octahedral space ${\cal O}$.
}
\end{figure}

\begin{figure}
\begin{center}
\vspace*{-90pt}
\begin{minipage}{11cm}
\hspace*{-65pt}\includegraphics[width=11.0cm]{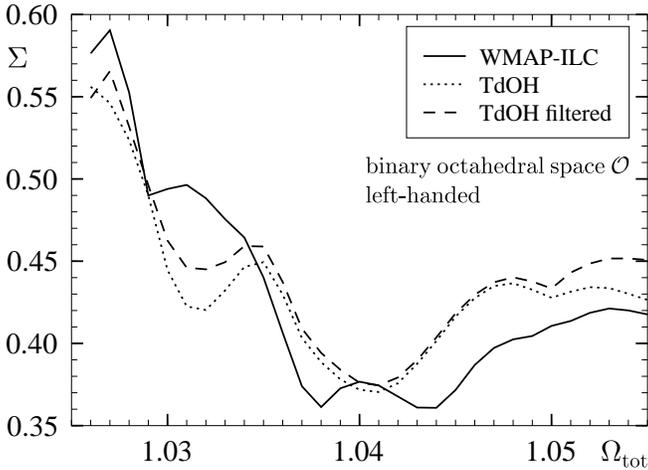}
\end{minipage}
\end{center}
\vspace*{-50pt}
\caption{\label{Fig:Sigma_WMAP_TdOH_Oktaeder_lh}
The same as in Fig.~\ref{Fig:Sigma_WMAP_TdOH_Dodekaeder_rh}
is shown for the left-handed binary octahedral space ${\cal O}$.
}
\end{figure}

The $\Sigma$ estimator is shown for the right- and left-handed
binary octahedral spaces ${\cal O}$ in
Figs.~\ref{Fig:Sigma_WMAP_TdOH_Oktaeder_rh} and
\ref{Fig:Sigma_WMAP_TdOH_Oktaeder_lh}, respectively.
For this space form, the best match to the WMAP angular power spectrum
is obtained for values around $\Omega_{\hbox{\scriptsize tot}}=1.038$
\citep{Aurich_Lustig_Steiner_2005a}.
The right-handed space displays only the decrease due to the
increasing size of the circles.
The left-handed case displays a minimum near
$\Omega_{\hbox{\scriptsize tot}}=1.04$ and $\Sigma$ increases
for higher values of $\Omega_{\hbox{\scriptsize tot}}$
only to values typical of the ${\cal S}^3$ simulation.

Figs.~\ref{Fig:Sigma_WMAP_TdOH_Tetraeder_rh} and
\ref{Fig:Sigma_WMAP_TdOH_Tetraeder_lh} show the $\Sigma$ estimator
for the tetrahedral space ${\cal T}$ again for the right- and
the left-handed variant, respectively.
In \citet{Aurich_Lustig_Steiner_2005a} it is found that the
suppression of large scale power agrees with the WMAP measurement
for $\Omega_{\hbox{\scriptsize tot}}$ in the range $1.06\dots 1.07$.
The large peak observed around $\Omega_{\hbox{\scriptsize tot}} = 1.056$
is caused by the increase of the number of circle pairs
which jumps from 4 to 7 (see Table \ref{paired_circles_tetrahedron}).
The right-handed space form displays large values around
$\Omega_{\hbox{\scriptsize tot}}\simeq 1.068$.
As in the case of the dodecahedron, the maximum is far less
pronounced than in the simulation of a ${\cal T}$ sky
as revealed by Fig.~\ref{Fig:Sigma_Tetraeder_Simulation},
and the remarks concerning the dodecahedron's maximum
applies also to this case.

\section{Conclusion}

We performed a thorough search for a circle-in-the-sky signature
in CMB maps under the hypothesis that the topology of the universe
is given by one of the 3 spherical spaces known as ${\cal D}$,
${\cal T}$, and ${\cal O}$.
To this purpose, we introduced the $\Sigma$ estimator (\ref{sigma_n_statistik})
which is applied to filtered sky maps using
the weight function (\ref{Eq:weight_function})
in order to enhance the ordinary Sachs-Wolfe contribution.
A possible signal was found in the case of the right-handed
Poincar\'e dodecahedral space ${\cal D}$ for
$\Omega_{\hbox{\scriptsize tot}} \simeq 1.015$,
see Fig.\ \ref{Fig:Sigma_WMAP_TdOH_Dodekaeder_rh},
which can be interpreted as a hint that our universe possesses
the topology of ${\cal D}$.
A similar possible signal occurred in the case of
the right-handed binary tetrahedral space around
$\Omega_{\hbox{\scriptsize tot}}\simeq 1.068$,
see Fig.\ \ref{Fig:Sigma_WMAP_TdOH_Tetraeder_rh}.
However, one must keep in mind that all three sky maps,
derived from the WMAP measurements in five frequency bands,
are a patchwork and thus possess a complicated signal to noise
behaviour.
Since the use of these incomplete sky maps certainly degrades a possible
topological signal, and in order to avoid a ``false negative'',
we can only report our findings without drawing a final conclusion
about the true topology of the Universe.
Furthermore, we believe that \citet{Cornish_Spergel_Starkman_Komatsu_2003}
overstate their result of not finding paired circles
since they also rely on the same CMB data.
The question whether our Universe possesses a non-trivial topology
is to our opinion not yet decided.


\bibliography{../bib_chaos,../bib_astro}
\bibliographystyle{mn2e}

\label{lastpage}

\end{document}